\documentclass[a4paper]{article}

\usepackage{INTERSPEECH2022}

\title{On incorporating social speaker characteristics in synthetic speech}
\name{Sai Sirisha Rallabandi$^1$, Sebastian Möller$^1$$^,$$^2$}
\address{
  $^1$Quality and Usability Lab, Technische Universität Berlin, Germany,\\ $^2$Speech and Language Technology, Deutsches Forschungszentrum für Künstliche Intelligenz (DFKI), Berlin, Germany}
\email{\{s.rallabandi, sebastian.moeller\}@tu-berlin.de}

\begin{document}

\maketitle
\begin{abstract}
In our previous work, we derived the acoustic features, that contribute to the perception of warmth and competence in synthetic speech. As an extension, in our current work, we investigate the impact of the derived vocal features in the generation of the desired characteristics. The acoustic features, spectral flux, F1 mean and F2 mean and their convex combinations were explored for the generation of higher warmth in female speech. The voiced slope, spectral flux, and their convex combinations were investigated for the generation of higher competence in female speech. We have employed a feature quantization approach in the traditional end-to-end tacotron based speech synthesis model. The listening tests have shown that the convex combination of acoustic features displays higher Mean Opinion Scores of warmth and competence when compared to that of individual features.


\end{abstract}

\noindent\textbf{Index Terms}: Social speaker characteristics, Warmth, Competence, Text-to-Speech synthesis, Tacotron

\section{Introduction}

Text-to-Speech synthesis (TTS) has evolved so much in the recent past \cite{zen16_interspeech, vandenoord16_ssw, DBLP:journals/corr/abs-1803-09017, Wang_2017, pmlr-v70-arik17a}. Current end-to-end speech synthesis systems have enabled various modifications in the synthesis procedure. Prosody modeling has been widely studied for the generation of expressivity, personality and various emotions in the generated speech. In \cite{DBLP:journals/corr/abs-1803-09017} style tokens have been explored to emphasize different parts of a sentence. This was further utilized for emotional speech generation in \cite{9023186}. In \cite{Tan2021FineGrainedSM}, authors employ a fine-grained style modeling by extracting the style information for expressive speech synthesis. Additionally, feature qunatization techniques have also demonstrated effective f0 modeling in TTS voices \cite{https://doi.org/10.48550/arxiv.2005.07884,8341752}.

TTS systems have also gained much interest in various applications \cite{DBLP:conf/interspeech/RaghaviRSB17,langlearning,Wilkinson2016OpenSourceCI,googleduplex}. The evaluation of these systems have consistently suggested improvements in the existing synthesis procedure \cite{govender18_interspeech, Anumanchipalli2010ImprovingSS, chang-2011-evaluation, evaltts}. In our previous work, we have studied the commercial TTS systems, Google \footnote{https://cloud.google.com/text-to-speech/} and Amazon voices \footnote{https://aws.amazon.com/polly/} \cite{rallabandi21_interspeech, rallabandi21_ssw}. Our study shows various speaker attributes contributing to the perception of the universal dimensions (warmth and competence) in synthetic speech \cite{rallabandi21_interspeech}. In \cite{rallabandi21_ssw}, we have derived the acoustic features that could affect warmth and competence in commercial TTS voices using linear regression. The speaker attributes we have considered in the study were as follows: friendliness and likability for warmth; skilfulness for competence. We have observed that the acoustic features, spectral flux, F1 mean, F2 mean, F3 mean are responsible for the speaker attribute, friendliness in female speech. The vocal features, spectral flux, f1 mean, f2 mean and voiced slope in the range of 500-1500 are responsible for likeability in female speech. The vocal features, voiced slope in the range of 0-500, spectral flux, mfcc contribute to speaker attribute, skilfulness in female voices. In our current work, we are interested in generating highly warm and highly competent female synthetic speech.

This paper is organised as follows: In Section \ref{sec:sysdesc}, we describe the system description followed by the details of the evaluation of the experiments in Section \ref{sec:eval}. In Section \ref{sec:disc} we provide a discussion on the study, followed by conclusions and future work in Section \ref{sec:Conclusions}. 

\section{System Description}
\label{sec:sysdesc}
\subsection{Baseline}
We run a traditional end-to-end Tacotron on LJSpeech as a baseline model \cite{Wang_2017, ljspeech17}. The train and test data sets are divided as 90\% and 10\% respectively. The tacotron model is fed with the phoneme sequence and the speech data. The phoneme sequence corresponding to the text was extracted from the FESTIVAL TTS \cite{Taylor1998TheAO}. For our current studies, we have utilized Griffin Lim algorithm for speech generation. The MOS scores obtained with the baseline on LJSpeech was 3.9.  

\subsection{Overview}

\begin{figure}[h]
  \centering
  \includegraphics[width=\linewidth]{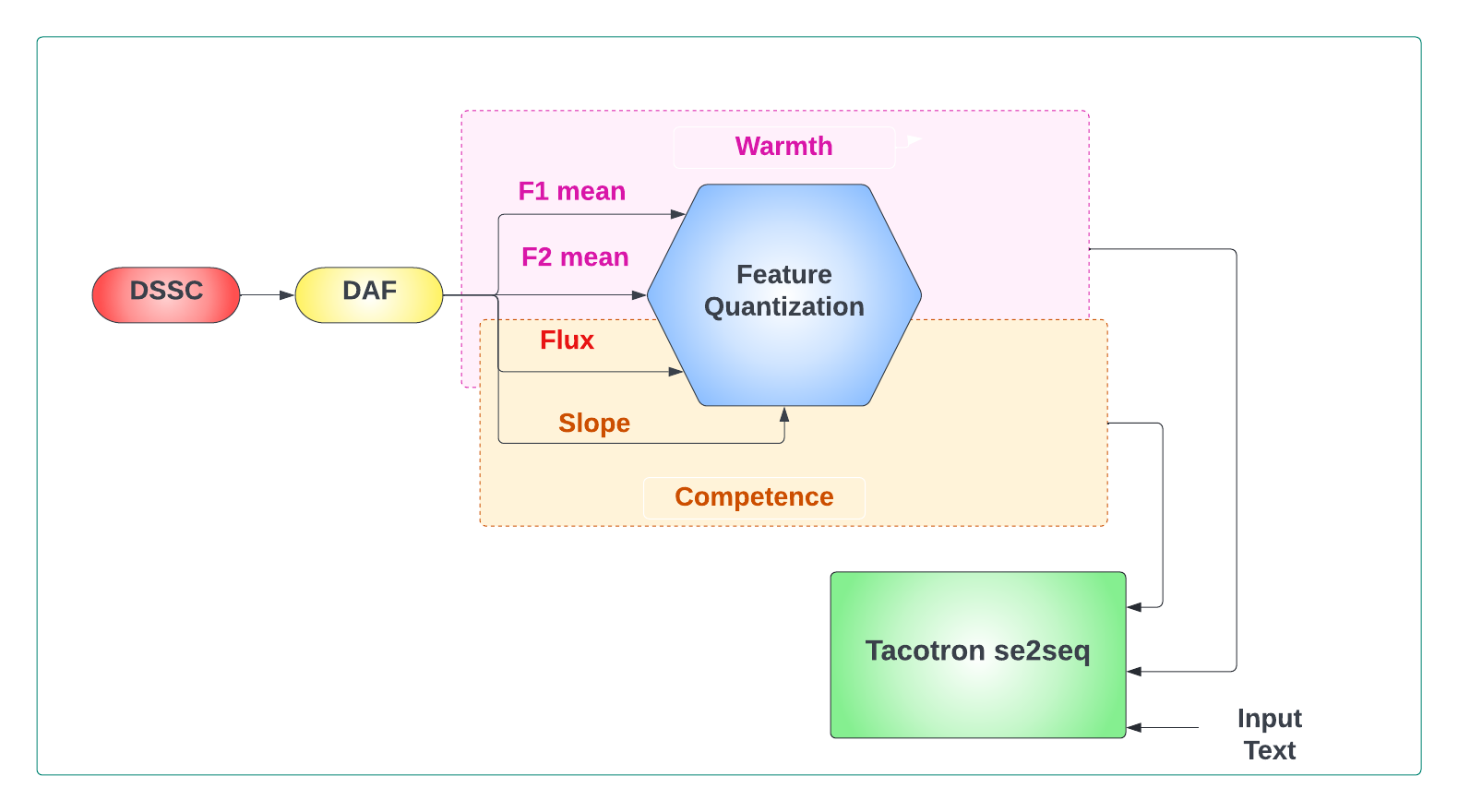}
  \caption{A schematic diagram of the current work. DSSC represents desired social speaker characteristics, DAF stands for derived acoustic features. Features corresponding to warmth and competence were quantized respectively and fed to the Tacotron model.}
  \label{fig:acoustic_TTS}
\end{figure}

Figure \ref{fig:acoustic_TTS} displays the block diagram of the current work. DSSC represents the desired social speaker characteristics from synthetic speech \cite{rallabandi21_interspeech}. DAV represents derived acoustic features \cite{rallabandi21_ssw}. To generate highly warm female speech, we studied the acoustic features that are commonly found in the speaker attributes, likeability and friendliness: F1 mean, F2 mean, spectral flux. Similarly, for competence, we investigated the features, spectral flux and voiced slope.

\subsection{Feature Quantization}

We have employed feature quantization on each of the acoustic features and have derived 3 different classes for each feature. Each class information was fed to the tacotron model as an additional dimension. Therefore, we perform feature dependent training for each of warmth and competence. The experimental details such as division of classes, examples per each class, class description are provided in the section \ref{sec:exp}

\subsection{TTS Experiments for Warmth}
\label{sec:exp}
We derived the openSMILE features on the LJspeech corpus as we were interested in various speaker characteristics \cite{opensmile}. We have extracted 88 acoustic features using the the Geneva Minimalistic Acoustic Parameter Set (eGeMAPS) configuration \cite{eyben2015geneva}.

\subsubsection{Experiment 1: F1 mean}
\begin{figure}[h]
  \centering
  \includegraphics[width=\linewidth]{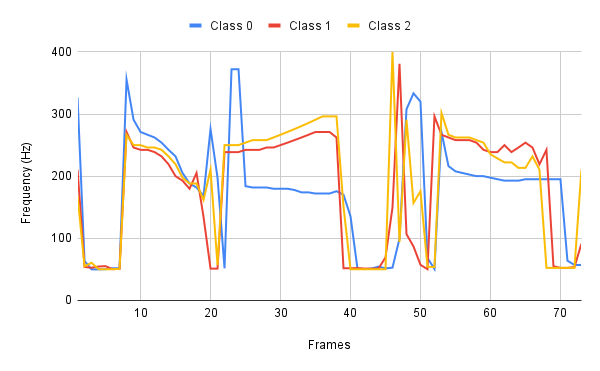}
  \caption{F0 contour of the generated speech corresponding to each class when trained with respect to F1 mean. The range of F1 mean values for each class are as follows, class 0 = 400 to 515, class 1 = 516 to 540, class 2 = 540 to 730.}
  \label{fig:F1mean_warmth}
\end{figure}

The openSMILE feature, F1 mean which is termed as, 'F1frequencysma3nzamean' was considered as one of the acoustic features responsible for both friendliness and likeability in female speech \cite{rallabandi21_ssw}. The maximum and minimum values were: 720.7 and 409.9 respectively. The feature quantization enabled 3 different classes (class 0, class 1, class 2) of F1 mean. The number of examples in each class were 4701, 3598 and 4801 respectively. The classes were termed as follows: class 0 = less warmth/cold, class 1 = neutral, class 2 = highest warmth. Figure \ref{fig:F1mean_warmth} displays the F0 contour of a speech segment. The variations in the F0 contour for all the 3 classes of the generated speech when conditioned on F1 mean are depicted. 

\subsubsection{Experiment 2: F2 mean}

\begin{figure}[h]
  \centering
  \includegraphics[width=\linewidth]{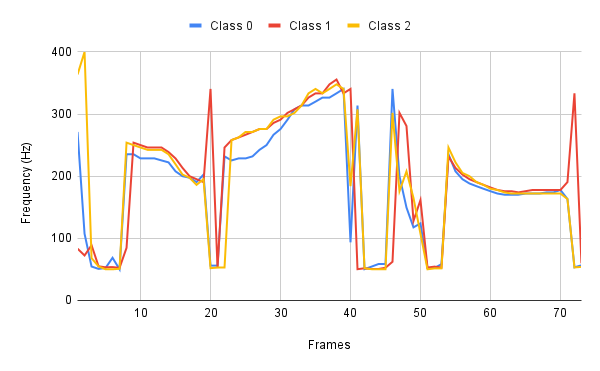}
  \caption{F0 contour of the generated speech corresponding to each class when trained with respect to F2 mean. The range of F2 mean values for each class are as follows, class 0 = 1280 to 1550, class 1 = 1551 to 1600, class 2 = 1601 to 1960.}
  \label{fig:F2mean_warmth}
\end{figure}

The openSMILE feature, F2 mean which is termed as, 'F2frequencysma3nzamean'
was considered as one of the acoustic features responsible for the speaker attributes friendliness and likeability in female speech \cite{rallabandi21_ssw}. The maximum and minimum values were: 1957.4 and 1280.5 respectively. The feature quantization enabled 3 different classes (class 0, class 1, class 2) of F2 mean. The number of examples in each class were 3410, 4431 and 5259 respectively. The classes were termed as follows: class 0 = less warmth/cold, class 1 = neutral, class 2 = highest warmth.
Figure \ref{fig:F2mean_warmth} displays the F0 contour of a speech segment. The variations in the F0 contour for all the 3 classes of the generated speech when conditioned on F2 mean are depicted. 

\subsubsection{Experiment 3: Spectral Flux}

\begin{figure}[h]
  \centering
  \includegraphics[width=\linewidth]{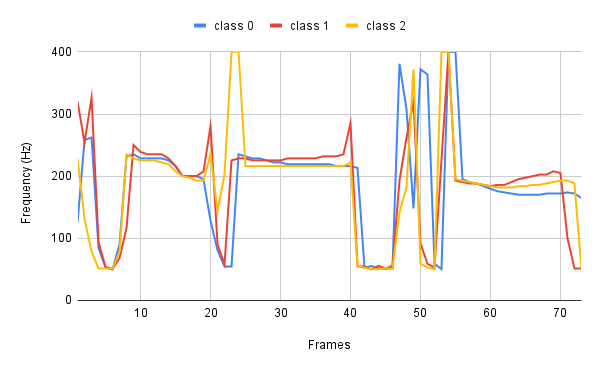}
  \caption{F0 contour of the generated speech corresponding to each class when trained with respect to Spectral Flux. The range of Spectral Flux values for each class are as follows, class 0 = 0 to 0.3, class 1 = 0.3 to 0.44, class 2 = 0.44 to 0.8.}
  \label{fig:Spectralflux_warmth}
\end{figure}

The openSMILE feature, Spectral Flux which is termed as, 'Spectralfluxsma3nzamean' was considered as one of the acoustic feature responsible for the speaker attributes friendliness and likeability in female speech \cite{rallabandi21_ssw}. The maximum and minimum values were: 0.706 and 0.15 respectively. The feature quantization enabled 3 different classes (class 0, class 1, class 2) of spectral flux. The number of examples in each class were 3193, 5193 and 4714 respectively. The classes were termed as follows: class 0 = less warmth/cold, class 1 = neutral, class 2 = highest warmth. Figure \ref{fig:Spectralflux_warmth} displays the F0 contour of a speech segment. The variations in the F0 contour for all the 3 classes of the generated speech when conditioned on Spectral Flux are depicted.

\subsubsection{Experiment 4: F1 mean + F2 mean + Spectral Flux}
\begin{figure}[h]
  \centering
  \includegraphics[width=\linewidth]{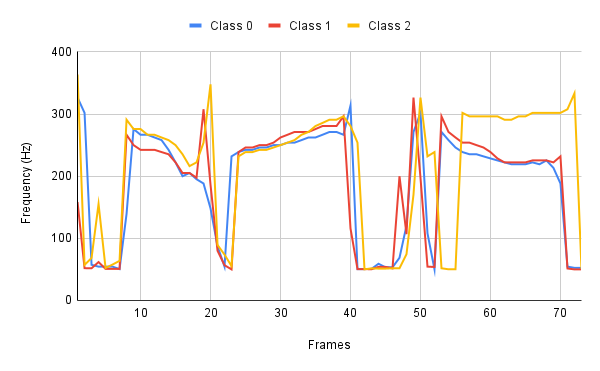}
  \caption{F0 contour of the generated speech corresponding to each class when trained with respect to convex combination of F1 mean, F2 mean and Spectral Flux. The range of values for each class are as follows, class 0 = 560 to 690, class 1 = 691 to 715, class 2 = 715 to 880.}
  \label{fig:warmthcxcomb}
\end{figure}

We have computed a convex combination of the vocal features, F1 mean, F2 mean and spectral flux. The maximum and minimum values were: 878.36 and 569.96 respectively. The feature quantization enabled 3 different classes (class 0, class 1, class 2) of convex combination. The number of examples in each class were 5073, 4458 and 3569 respectively. The classes were termed as follows: class 0 = less warmth/cold, class 1 = neutral, class 2 = highest warmth. Figure \ref{fig:warmthcxcomb} displays the F0 contour of a speech segment. The variations in the F0 contour for all the 3 classes of the generated speech when conditioned on the convex combination of F1 mean, F2 mean and Spectral Flux are depicted. 

\subsection{TTS Experiments for Competence}
\subsubsection{Experiment 5: Voiced Slope}

\begin{figure}[h]
  \centering
  \includegraphics[width=\linewidth]{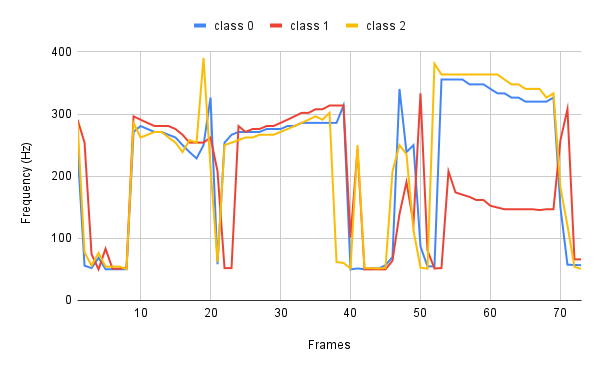}
  \caption{F0 contour of the generated speech corresponding to each class when trained with respect to Voiced slope. The range of slope values for each class are as follows, class 0 = 0.07 to 0.11, class 1 = 0.112 to 0.116, class 2 = 0.116 to 0.139.}
  \label{fig:slope_comp}
\end{figure}

Slope was considered as one of the acoustic feature responsible for the speaker attribute, skilfulness in female speech \cite{rallabandi21_ssw}. The maximum and minimum values were: 0.139 and 0.072 respectively. The feature quantization enabled 3 different classes (class 0, class 1, class 2) of voiced slope. The number of examples in each class were 4701, 3598 and 4801 respectively. The classes were termed as follows: class 0 = less warmth/cold, class 1 = neutral, class 2 = highest warmth. Figure \ref{fig:slope_comp} displays the F0 contour of a speech segment. The variations in the F0 contour for all the 3 classes of the generated speech when conditioned on slope are depicted.

\subsubsection{Experiment 6: Voiced Slope + Spectral Flux}

\begin{figure}[h]
  \centering
  \includegraphics[width=\linewidth]{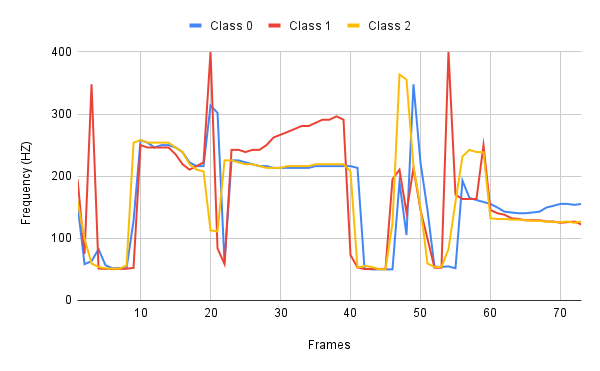}
  \caption{F0 contour of the generated speech corresponding to each class when trained with respect to Convex combination of Spectral Flux and Voiced Slope. The range of values for each class are as follows, class 0 = 0.133 to 0.22, class 1 = 0.23 to 0.26, class 2 = 0.27 to 0.45.}
  \label{fig:convx_comb_comp}
\end{figure}

The convex combinations of slope and spectral flux were considered in this experiment The maximum and minimum values were: 0.411 and 0.133 respectively. The feature quantization enabled 3 different classes (class 0, class 1, class 2) of voiced slope. The number of examples in each class were 4882, 4365 and 3853 respectively. The classes were termed as follows: class 0 = incompetent, class 1 = neutral, class 2 = highly competent. Figure \ref{fig:convx_comb_comp} displays the F0 contour of a speech segment. The variations in the F0 contour for all the 3 classes of the generated speech when conditioned on the convex combination of slope and spectral Flux are depicted.

\section{Evaluation}
\label{sec:eval}
The subjective evaluation was conducted for all the experiments performed for warmth and competence. As shown in \cite{rallabandi21_ssw}, warmth was examined on the scales, likeability and friendliness. Correspondingly, competence was examined on the scale skilfulness.  We have recruited 25 University students for our listening tests. We provide a 5-point likert scale during the listening tests where, 5 = highly friendly/likeable/skilfull, 1 = unfriendly/unlikable/unskillfull. We have provided 15 sentences (15 = 5 sentences *3 classes per sentence) for each of F1 mean, F2 mean, spectral Flux, Slope and convex combinations of F1 mean, F2 mean, flux and slope for warmth. Thus, our listening test consisted of 90 sentences. For competence, the sentences generated from experiments, 5 and 6 were provided (10 sentences * 3 classes per each of slope and convex combination). The listeners could listen to the speech samples any number of times during the listening test. The sentences were randomized for each participant. 

The sentences provided in the listening test are presented below.

\begin{itemize}
    \item \textit{Suggestions for improvement means a person believes in your core idea and thinks their comments will help your work.}
    \item \textit{Don’t put time on it. Relax! May be nap and get back to it when you get up.}
    \item \textit{Don’t be disheartened, that’s normal. It’s part of the process.}
    \item \textit{Maybe the lesson here is that it's very hard to have a totally relaxed interpersonal relationship!}
    \item \textit{I have been exactly here. I am always ready if you ever need support. I'm here for you!}
\end{itemize}

\begin{figure}[h]
  \centering
  \includegraphics[width=\linewidth]{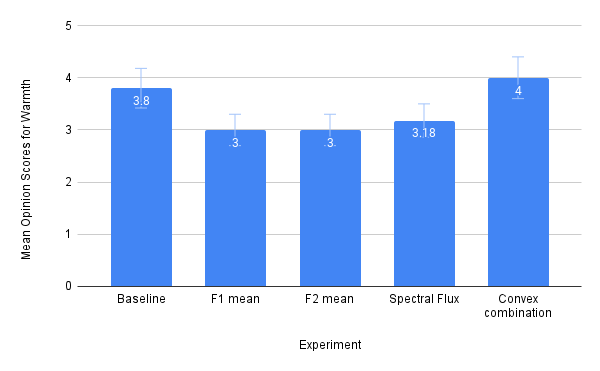}
  \caption{MOS scores of warmth.}
  \label{fig:warmthboxplots}
\end{figure}

Figure \ref{fig:warmthboxplots} displays the plot with the Mean Opinion scores (MOS) collected for the characteristic, warmth across baseline, the acoustic features, F1 mean, F2 mean, Spectral Flux and their convex combinations. In order to obtain the MOS scores for warmth, we have averaged the subjective ratings of friendliness and likability. We also provided the error bars for each experiment. The Mean Opinion Scores (MOS) are obtained as follows: baseline = 3.8, F1 mean = 3, F2 mean = 3, Spectral Flux = 3.18, Convex combination of F1 mean, F2 mean, and Spectral Flux = 4. We observed that the MOS scores obtained for convex combination of the acoustic features resulted in higher warmth when compared to that of individual acoustic features and the baseline. The generated speech when conditioned on F1 mean and F2 mean displayed similar MOS scores. The Spectral Flux displayed slightly higher warmth that that of F1 mean and F2 mean but lower than that of the baseline.

\begin{figure}[h]
  \centering
  \includegraphics[width=\linewidth]{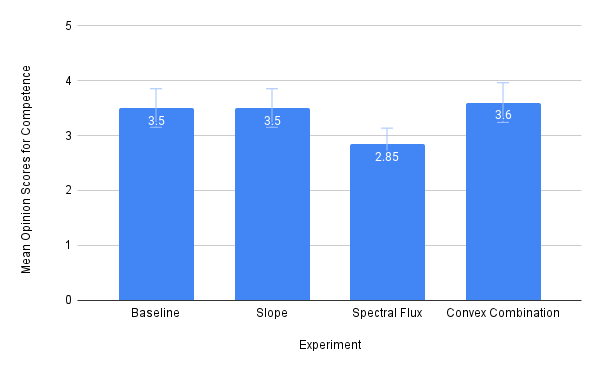}
  \caption{MOS ratings of competence.}
  \label{fig:compboxplots}
\end{figure}

Figure \ref{fig:compboxplots} displays the plot with the Mean Opinion scores (MOS) collected for the characteristic, competence across the acoustic features, Slope, Spectral Flux and their convex combinations. In order to obtain the MOS scores for competence, we have considered the subjective ratings of skilfulness. We also provided the error bars for each experiment. The scores are obtained as follows: baseline = 3.5, Slope = 3.5, Spectral Flux = 2.85, Convex combination of Slope, and Spectral Flux = 3.6. We observed that the MOS scores obtained for convex combination of the acoustic features resulted in higher competence when compared to that of individual acoustic features and the baseline. The generated speech with the baseline and when conditioned on Slope displayed similar MOS scores on competence. The Spectral Flux displayed lowest competence scores among all the experiments. 

\section{Discussion}
\label{sec:disc}
In the current study, our experiments were conducted on LJSpeech database. As the content used during training the TTS was non-fiction passages, we assume that the models trained on read speech or other datasets might display a different set of results. Also, we assume that the feature quantization we have employed might be specific to our study. Furthermore, the sentences chosen for our subjective evaluation display compassion. Therefore, we assume that the content of the sentences would have had some impact on the subjective ratings. Our previous studies were conducted on synthetic speech. In the current work, we chose to quantize the same acoustic features in natural speech. The hypothesis was that, since the conditioning of the acoustic features was done during the speech generation, using the same features for feature quantization on natural speech should provide similar results. 


\section{Conclusions and Future work}
\label{sec:Conclusions}
This paper is an extension of our previous work, where we have derived the acoustic features contributing to the perception of warmth and competence in female synthetic speech. The characteristic, warmth was studied through the speaker attributes, likability and friendliness. The characteristic, competence was studied through the speaker attribute, skilfulness. We have employed feature quantized acoustic feature dependent training using a traditional Tacotron model. The listening tests have manifested that the convex combination of the acoustic features renders higher warmth and competence in synthetic speech than individual speech features. Additionally, we found that the speech generated when conditioned on individual acoustic features displayed lower or equal MOS scores of warmth and competence with that of the baseline. The future work could be to investigate the feature quantization on male synthetic speech. Additionally, a comparison between the female and male speech could also be an interesting study.


\section{Acknowledgements}

Authors would like to thank Sai Krishna Rallabandi for his valuable time and feedback. This work is being supported by the German Research Foundation (DFG), under funding MO 1038/29-1, TU PSP-Element: 1-50001062-01-EF. We also thank all the participants of our subjective tests.

\bibliographystyle{IEEEtran}

\bibliography{mybib}

\end{document}